\documentclass{article}

\usepackage{arxiv}

\usepackage[utf8]{inputenc} % allow utf-8 input
\usepackage[T1]{fontenc}    % use 8-bit T1 fonts
\usepackage{hyperref}       % hyperlinks
\usepackage{url}            % simple URL typesetting
\usepackage{booktabs}       % professional-quality tables
\usepackage{amsfonts}       % blackboard math symbols
\usepackage{nicefrac}       % compact symbols for 1/2, etc.
\usepackage{microtype}      % microtypography
\usepackage{lipsum}		% Can be removed after putting your text content
\usepackage{graphicx}

\usepackage{amsmath}
\usepackage{chemformula}

\usepackage{epstopdf}
\usepackage{array}

\title{Low-temperature plasma simulation based on physics-informed neural networks: \\frameworks and preliminary applications}

\author{ Linlin Zhong\thanks{This paper is currently under consideration by a journal. The author's homepage:\href{https://mathboylinlin.com}{https://mathboylinlin.com}}, ~Bingyu Wu, and Yifan Wang \\
	School of Electrical Engineering\\
	Southeast University\\
	No 2 Sipailou, Nanjing, Jiangsu Province 210096, P. R. China\\
	\texttt{mathboylinlin@gmail.com}, \texttt{linlin@seu.edu.cn}\\
}

\date{June 30, 2022}

\renewcommand{\headeright}{}
\renewcommand{\undertitle}{A Preprint}

\begin{document}
\maketitle

\begin{abstract}
Plasma simulation is an important and sometimes only approach to investigating plasma behavior. In this work, we propose two general AI-driven frameworks for low-temperature plasma simulation: Coefficient-Subnet Physics-Informed Neural Network (CS-PINN) and Runge-Kutta Physics-Informed Neural Network (RK-PINN). The CS-PINN uses either a neural network or an interpolation function (e.g. spline function) as the subnet to approximate solution-dependent coefficients (e.g. electron-impact cross sections, thermodynamic properties, transport coefficients, et al.) in plasma equations. On the basis of this, the RK-PINN incorporates the implicit Runge-Kutta formalism in neural networks to achieve a large-time-step prediction of transient plasmas. Both CS-PINN and RK-PINN learn the complex non-linear relationship mapping from spatio-temporal space to equation’s solution. Based on these two frameworks, we demonstrate preliminary applications by four cases covering plasma kinetic and fluid modeling. The results verify that both CS-PINN and RK-PINN have good performance in solving plasma equations. Moreover, the RK-PINN has ability of yielding a good solution for transient plasma simulation with not only large time step but also limited noisy sensing data. 
\end{abstract}

\section{INTRODUCTION}
\paragraph{}
Plasma is an electrically quasi-neutral medium consisting of electrons, ions, and neutral atoms or molecules in a gaseous state. It has a wide range of industrial applications, especially for low-temperature plasmas, such as electronic manufacturing  \cite{zhang2018amoo3}, material synthesis \cite{han2018recent}, arc interruption \cite{zhong2019animproved}, combustion ignition \cite{li2021advances}, space propulsion  \cite{Levchenko2020perspectives}, waste treatment \cite{sanito2021application}, and even biological medicine \cite{cheng2020onthe}. To well understand plasma behavior in the aforementioned applications, many experimental and computational methods have been developed over the past few decades \cite{loureiro2016kinetics}. However, due to the complex multi-physics phenomena in plasmas and limited measurement tools under extreme operating conditions, plasma simulation is sometimes the only approach to investigating plasmas and quantifying their characterization. For instance, chemical kinetic modeling is always used in plasma chemistry, which can provide more detailed quantitative information than any experiments about chemical reactions associated with plasma species \cite{zhong2021dynamics}. Another example, in arc plasmas, the challenges of measuring multi-physics fields arising from high plasma temperatures can be overcome by fluid modeling, e.g. Magnetohydrodynamics (MHD) modeling \cite{gleizes2014perspectives}. In essence, both kinetic and fluid simulation for plasmas is mathematically solving partial differential equations (PDEs) with defined initial and boundary conditions in a given spatio-temporal domain. The numerical methods including finite element method (FEM) and finite volume method (FVM), are commonly used to solve plasma governing equations. Typically, both FEM and FVM require sophisticated meshing on a computational domain, which is not an easy job for complex irregular geometry, particularly in the modeling of practical applications. Poor meshing also leads to poor numerical solution and even convergence problem. Worse still, in order to capture some subtle features of plasma dynamics, such as small eddies and shock waves, fine grids rather than coarse grids are necessarily needed, which however means heavy and even infeasible computation cost \cite{kochkov2021machine}.

\paragraph{}
Recently, with the development of artificial intelligence (AI), a few AI-driven methods emerge as a new technique for numerically solving PDEs. They make best use of the powerful capability of neural networks to represent non-linear relationships in PDEs. The most typical representative is the Physics-Informed Neural Networks (PINNs) proposed by Karniadakis and his students \cite{raissi2019physics}. In PINNs, the physics constraints described by PDEs are directly incorporated in the loss function of neural networks, and the solution of differential equations is approximated by minimizing the corresponding loss function \cite{cai2021physicsfluid}. Compared with FEM and FVM, PINNs are a mesh-free method which has high flexibility and can easily adapt to complex-geometry domain \cite{pang2019fpinns}. It has been widely used in solving various classical PDEs since proposed \cite{raissi2019physics, pang2019fpinns, jagtap2020extended, chen2021learning}, and its improved variants are springing up like wildfire inspired by the latest deep learning techniques e.g. generative adversarial network (GAN) \cite{yang2019adversarial}, attention mechanism \cite{rodriguez2022physics}, and adaptive activation function \cite{jagtap2020adptive}. In the field of scientific computing, PINNs are also applied to solve traditional physics and chemistry problems, such as fluid dynamics \cite{raissi2020hidden}, heat transfer \cite{cai2021physicsheat}, atomic interaction \cite{pun2019physically}, and chemical kinetics \cite{ji2021stiff}. However, perhaps due in part to the very complex multi-physics coupling relationships in plasmas, there are few works on the PINN-based plasma modeling, except for the works by Kawaguchi et al. \cite{kawaguchi2020deep} for Boltzmann equation of electrons and our previous work on the thermal plasma simulation \cite{zhong2020deep}. Considering that the computing paradigm represented by PINNs could have potential to bring a revolution in solving PDEs, it is very interesting and necessary to construct a PINN-based framework for plasma simulation and explore the corresponding feasibility and generality. 
 
\paragraph{} 
Consequently, this work proposes two general AI-driven simulation frameworks for low-temperature plasmas based on PINNs and demonstrates their preliminary applications by four cases covering plasma kinetic and fluid modeling. The rest of the paper is organized as follows. In Section II, the two proposed PINN-based frameworks, namely the Coefficient-Subnet Physics-Informed Neural Network (CS-PINN) and Runge-Kutta Physics-Informed Neural Network (RK-PINN), are described in detail. In Section III, the four cases including the plasma equations governing electron distribution, corona discharge and arc evolution, are studied based on CS-PINN and/or RK-PINN. Finally, we conclude this work in Section IV.

\section{FRAMEWORKS OF AI-DRIVEN PLASMA SIMULATION}
\label{sec:sec2}
\paragraph{}
Typically, there are two approaches to simulating plasma behavior: kinetic and fluid modeling. The former approach is to describe plasmas from a microscopic point of view by solving a kinetic equation for distribution functions of plasma particles, whereas the latter approach is to describe plasmas as continuous fluid from a macroscopic view. In some cases, the kinetic and fluid modeling are used together in a hybrid way. Whichever model is used, a group of PDEs (e.g. the Boltzmann equation for kinetic model \cite{hagelaar2005solving} and the drift-diffusion equations for fluid model \cite{gao2018numerical}) have to be solved in a proper way. This means the idea of PINNs for solving PDEs can be used in plasma simulation. Meanwhile, the special features of plasma simulation must also be considered.

\subsection{Coefficient-Subnet Physics Informed Neural Network (CS-PINN) for solving plasma equations with solution-dependent coefficients}
\label{sec:sec2.1}
\paragraph{}
We consider the following general plasma equation defined on spatio-temporal domain $\Omega$ with boundary and initial conditions defined on $\partial\Omega$.

\begin{equation}
\label{equ:equ1}
	f\left(t, \mathbf{x}, \partial_{t} u, \partial_{\mathbf{x}} u, \cdots, \lambda\right)=0
\end{equation}

\begin{equation}
\label{equ:equ2}
	\mathcal{B}(u, t, \mathbf{x})=0, \mathcal{I}(u, \mathbf{x})=0
\end{equation}

Where $\mathbf{x} \in \mathbb{R}^{d}$ and $t$ are the spatial and temporal coordinates respectively; $u(t, \mathbf{x})$ is the solution and $\lambda=\left\lceil\lambda_{1}, \lambda_{2}, \cdots\right]$ are the coefficients in the governing equation. In some cases, $\lambda$ may also appear in initial and boundary conditions.

\begin{figure}
	\centering
	\includegraphics[width=9.5cm]{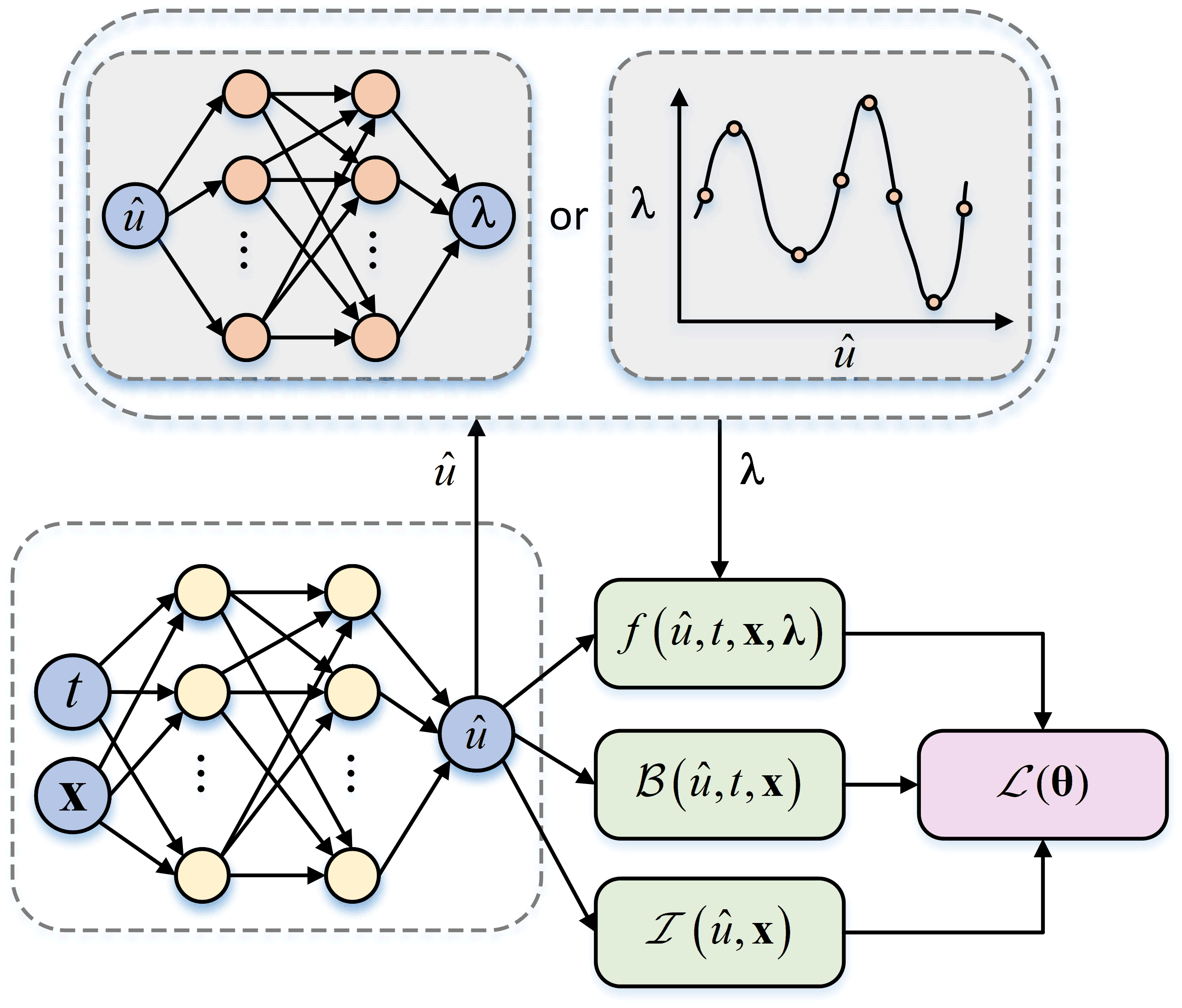}
	\caption{AI-driven plasma simulation framework: Coefficient-Subnet Physics Informed Neural Network (CS-PINN) for solving plasma equations with solution-dependent coefficients.}
	\label{fig:fig1}
\end{figure}

\paragraph{}
As shown in Figure \ref{fig:fig1}, we construct a neural network which takes $(t, \mathbf{x})$ as input and outputs $\hat{u}(t, \mathbf{x})$ as the surrogate of the equation’s solution $u(t, \mathbf{x})$. Based on the automic differentiation, any gonverning equations as well as initial and boundary conditions with any differential operators can be easily expressed by the neural network. The basic idea of PINNs is to find a well-trained neural network which satisfies the equations (\ref{equ:equ1}) and (\ref{equ:equ2}). This can be doned by incorporating the equations in the loss function and then minimizing it iteratively using a opitminzation algorithm e.g. stochastic gradient descent algorithm or its variants. Generally, the loss function of PINNs has the form as follows \cite{raissi2019physics, cai2021physicsfluid, zhong2020deep}).

\begin{equation}
\label{equ:equ3}
	\mathcal{L}=\omega_{f} \mathcal{L}_{f}+\omega_{\mathcal{B}} \mathcal{L}_{\mathcal{B}}+\omega_{\mathcal{I}} \mathcal{L}_{\mathcal{I}}
\end{equation}

\begin{equation}
\label{equ:equ4}
	\mathcal{L}_{f}=\frac{1}{N_{f}} \sum_{i=1}^{N_{f}} \mathcal{F}\left(f\left(t_{i}, \mathbf{x}_{i}, \partial_{t} \hat{u}_{i}, \partial_{\mathbf{X}} \hat{u}_{i}, \cdots, \lambda\right)\right)
\end{equation}

\begin{equation}
\label{equ:equ5}
	\mathcal{L}_{\mathcal{B}}=\frac{1}{N_{\mathcal{B}}} \sum_{i=1}^{N_{\mathcal{B}}} \mathcal{F}\left(\mathcal{B}\left(\hat{u}_{i}, t_{i}, \mathbf{x}_{i}\right)\right)
\end{equation}

\begin{equation}
\label{equ:equ6}
	\mathcal{L}_{\mathcal{I}}=\frac{1}{N_{\mathcal{I}}} \sum_{i=1}^{N_{\mathcal{I}}} \mathcal{F}\left(\mathcal{I}\left(\hat{u}_{i}, \mathbf{x}_{i}\right)\right)
\end{equation}

Where $\omega_{f}$, $\omega_{\mathcal{B}}$, and $\omega_{\mathcal{I}}$ are the weighting factors regulating different loss terms; $N_{f}$, $N_{\mathcal{B}}$, and $N_{\mathcal{I}}$ are the number of the scattered points we select uniformly or randomly in the computational domain $\Omega$ and the boundary and initial domain $\partial\Omega$. $\mathcal{F}(\cdot)$ is the function for measuring the residuals of equations. Usually, it can be $L^{1}$ or $L^{2}$ norm or their combination \cite{zhong2020deep}.

\paragraph{}
However, compared with general PDEs, the governing equations of plasmas have solution-dependent coefficients in themselves, which means the parameters $\lambda$ in the equations (\ref{equ:equ1}) are not constant but variational with the solution. For example, the Boltzmann equation describing weakly ionized plasmas has various electron-impact cross sections (e.g. elastic, momentum transfer, excitation, and ionization cross sections) which depend on incident electron energy. In fluid modeling, there are also many solution-dependent coefficients, such as mass density in continuity equation, viscosity in momentum equation, thermal conductivity in energy equation, and electrical conductivity in electric field equation. All these coefficients are usually precalculated and given in a tabular way as function of electron energy, temperature, pressure, and plasma composition. 

\paragraph{}
To effectively express the non-linear relationship mapping from equation’s solution to coefficients, we propose to add a subnet as the part of the PINN-based framework for plasma simulation, i.e. the Coefficient-Subnet Physics Informed Neural Network (CS-PINN). This subnet can be either a neural network or an interpolation function (e.g. spline function). Our previous work has verified that a small feed-forward neural network has sufficient capability to approximate thermodynamic and transport coefficients in thermal plasma equations \cite{zhong2020deep}. The disadvantage is that we have to pretrain the subnet for expressing coefficients before training the entire CS-PINN. This requires extra time consumption and sometimes causes overfitting in the nearby boundary. If it is not very critical to the smoothness of coefficient approximator, we can use interpolation functions as subnet instead. Considering that most plasma equations have no more than two-order differential operators, the coefficient approximator by cubic spline function is smooth enough to be used in the CS-PINN. Furthermore, large neural networks with many layers and neurons are sometimes needed to approximate the coefficients varying very significantly with solution, e.g. the temperature-dependent thermal conductivity in the energy conservation equation and the electron-energy-dependent excitation cross sections in the Boltzmann equation. This could result in a huge amount of computation and make the spline-function-based subnet superior over the neural-network-based subnet. It is notable that when there are high-order differential operators in plasma equations, neural networks are the better choice than interpolation functions as the subnet in CS-PINN.

\subsection{Runge-Kutta Physics Informed Neural Network (RK-PINN) for solving plasma equations with transient terms}
\label{sec:sec2.2}
\paragraph{}
In the PINN-based method (including CS-PINN), loss functions are calculated by the scattered data points selected over the entire spatio-temporal domain. This means a $N$-dimensional problem requires training the surrogate network in the $(N+1)$-dimensional data space. The extra dimension comes from the temporal space, and this undoubtedly increases the difficulty of model training. To avoid the data discretization along the temporal coordinate, Raissi et al. \cite{raissi2019physics} applied the Runge-Kutta method in the PINNs for solving simultaneous PDEs. In this work, we incorporate this formalism into the CS-PINN and yield the Runge-Kutta Physics Informed Neural Network (RK-PINN) for solving plasma equations with transient terms.

\paragraph{} 
Let’s consider the following two coupled equations, one of which has a transient term and the other not.

\begin{equation}
\label{equ:equ7}
	\frac{\partial u}{\partial t}=f\left(\mathbf{x}, \partial_{\mathbf{x}} u, \partial_{\mathbf{x}} v, \cdots, \lambda\right)
\end{equation}

\begin{equation}
\label{equ:equ8}
	g \left(\mathbf{x}, \partial_{\mathbf{x}} u, \partial_{\mathbf{x}} v, \cdots, \lambda\right)=0
\end{equation}

With the boundary condition

\begin{equation}
\label{equ:equ9}
	\mathcal{B}_{f}(u,v,\mathbf{x})=0,\mathcal{B}_{g}(u,v,\mathbf{x})=0
\end{equation}

\paragraph{} 
According to the classic Runge-Kutta method, the solution of a transient equation can be approximated by an implicit or explicit iterative method given an initial condition. To allow a large time step with stability, an implicit Runge-Kutta formalism with q stages is used in the RK-PINN as follows \cite{raissi2019physics}.

\begin{equation}
\label{equ:equ10}
	\left\{\begin{array}{l}u\left(t_{0}+\Delta t \cdot c_{1}\right)=u\left(t_{0}\right)+\Delta t \sum_{j=1}^{q} a_{i j} f\left(u\left(t_{0}+\Delta t \cdot c_{j}\right)\right) \\ u\left(t_{0}+\Delta t \cdot c_{2}\right)=u\left(t_{0}\right)+\Delta t \sum_{j=1}^{q} a_{i j} f\left(u\left(t_{0}+\Delta t \cdot c_{j}\right)\right) \\ \cdots \ldots \\ u\left(t_{0}+\Delta t \cdot c_{q}\right)=u\left(t_{0}\right)+\Delta t \sum_{j=1}^{q} a_{i j} f\left(u\left(t_{0}+\Delta t \cdot c_{j}\right)\right) \\ u\left(t_{0}+\Delta t\right)=u\left(t_{0}\right)+\Delta t \sum_{j=1}^{q} b_{j} f\left(u\left(t_{0}+\Delta t \cdot c_{j}\right)\right)\end{array}\right.
\end{equation}

Where $\left\{a_{i j}, b_{j}, c_{j}\right\}$ are the parameters in the implicit Runge-Kutta formalism, which are usually tabulated in a so-called Butcher table.

\begin{figure}
	\centering
	\includegraphics[width=9.5cm]{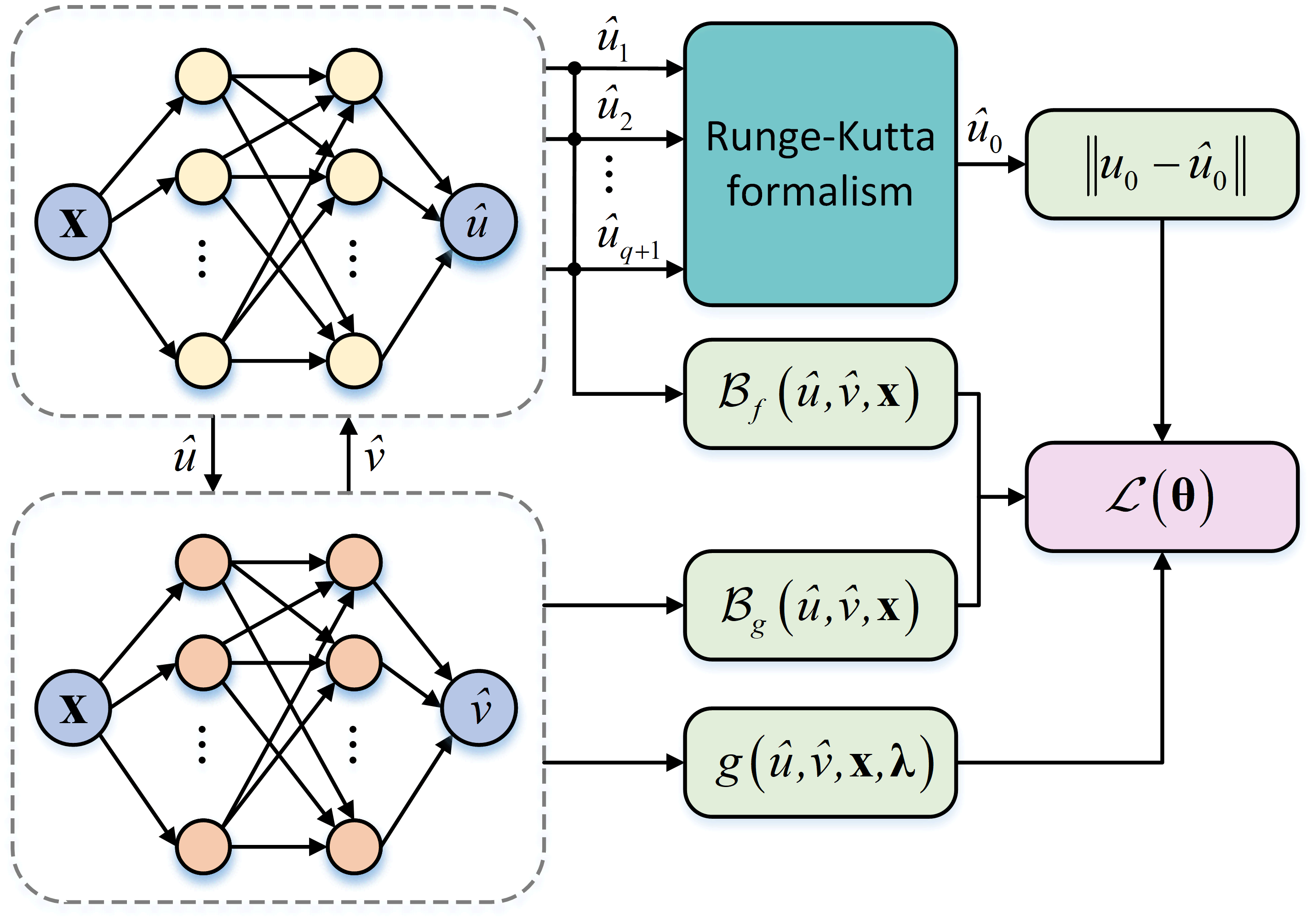}
	\caption{AI-driven plasma simulation framework: Runge-Kutta Physics Informed Neural Network (RK-PINN) for solving plasma equations with transient terms.}
	\label{fig:fig2}
\end{figure}

\paragraph{} 
As shown in Figure \ref{fig:fig2}, we construct two neural networks as the surrogates of equations  and  respectively. Both networks take $\mathbf{x}$ as input and output their each solutions $\hat{u}$ and $\hat{v}$. The difference is that the output of the former network consists of $q+1$ elements corresponding to the solutions at different steps of Runge-Kutta formalism. The solution-dependent coeffcients in both equations  and  can also be approximated by subnets as CS-PINN. The loss function is composed of four terms.

\begin{equation}
\label{equ:equ11}
	\mathcal{L}=\omega_{f} \mathcal{L}_{f}+\omega_{\mathcal{B}_{f}} \mathcal{L}_{\mathcal{B}_{f}}+\omega_{\mathcal{B}_{g}} \mathcal{L}_{\mathcal{B}_{g}}+\omega_{g} \mathcal{L}_{g}
\end{equation}

\begin{equation}
\label{equ:equ12}
	\mathcal{L}_{f}=\frac{1}{N_{f}} \sum_{i=1}^{N_{f}} \sum_{j=1}^{q+1}\left\|u_{0}^{i}-\hat{u}_{0}^{i, j}\right\|
\end{equation}

\begin{equation}
\label{equ:equ13}
	\mathcal{L}_{\mathcal{B}_{f}}=\frac{1}{N_{\mathcal{B}_{f}}} \sum_{i=1}^{N_{\mathcal{B}_{f}}} \mathcal{F}\left(\mathcal{B}_{f}\left(\hat{u}_{i}, \hat{v}_{i}, \mathbf{x}_{i}\right)\right)
\end{equation}

\begin{equation}
\label{equ:equ14}
	\mathcal{L}_{\mathcal{B}_{g}}=\frac{1}{N_{\mathcal{B}_{g}}} \sum_{i=1}^{N_{\mathcal{B}_{g}}} \mathcal{F}\left(\mathcal{B}_{g}\left(\hat{u}_{i}, \hat{v}_{i}, \mathbf{x}_{i}\right)\right)
\end{equation}

\begin{equation}
\label{equ:equ15}
	\mathcal{L}_{g}=\frac{1}{N_{g}} \sum_{i=1}^{N_{g}} \mathcal{F}\left(g\left(\mathbf{x}_{i}, \partial_{\mathbf{X}} \hat{u}_{i}, \partial_{\mathbf{X}} \hat{v}_{i}, \cdots, \lambda\right)\right)
\end{equation}

Where $\omega_{f}$, $\omega_{\mathcal{B}_{f}}$, $\omega_{\mathcal{B}_{g}}$ and $\omega_{g}$ are the weighting factors regulating different loss terms; $N_{f}$, $N_{\mathcal{B}_{f}}$, $N_{\mathcal{B}_{g}}$, and $N_{g}$ are the number of the scattered points in the spatial domain and its boundary; $u_{0}^{i}$ are the given initial values of the solution $u_{i}$ at the scattered points; $\hat{u}_{0}^{i, j}$ are the prediction of $u_{0}^{i}$ by the neural network, which are calculated as follows by moving the second term on the right side of the equations  to the left.

\begin{equation}
\label{equ:equ16}
	\hat{u}_{0}^{i, j}=\hat{u}\left(t_{0}+\Delta t \cdot c_{i}\right)-\Delta t \sum_{j=1}^{q} a_{i j} f\left(\hat{u}\left(t_{0}+\Delta t \cdot c_{j}\right)\right), \quad i=1,2, \cdots, q
\end{equation}

\begin{equation}
\label{equ:equ17}
	\hat{u}_{0}^{i, q+1}=\hat{u}\left(t_{0}+\Delta t\right)-\Delta t \sum_{j=1}^{q} b_{j} f\left(\hat{u}\left(t_{0}+\Delta t \cdot c_{j}\right)\right)
\end{equation}

\paragraph{} 
After the well training of the two coupled neural networks, it is achieved that the solutions $\hat{u}$ and $\hat{v}$ at the time step $\Delta t$ can be calculated directly from their intial values. The implicit Runge-Kutta formalism with larger stage $q$ allows us to use larger time step $\Delta t$ in the calculation. This means it could be possible to directly obtain a final evolutionary state of a plasma starting from an intial state after a given period even if this period is long. Moreover, compared with CS-PINN, the RK-PINN does not need to add extra dimension in the data sampling. Therefore, the RK-PINN consumes less computational memory in both training and inference, and it should be faster than the CS-PINN. It is worth noting that we can merge the two networks and yield a single network by adding extra nodes in the output layer and sharing the weights. This could be useful when the two equations have very strong coupling.

\section{CASE STUDY}
\label{sec:sec3}
\paragraph{}
To demonstrate the feasibility and versatility of the two proposed frameworks for plasma simulation, we study an argon plasma as example and solve the Boltzmann equation, the drift-diffusion-Poisson equations, and the time dependent Elenbaas-Heller equations based on CS-PINN and/or RK-PINN. We also demonstrate how to merge sensing data in the simulation when a part of equation conditions is unavailable. All the calculations are performed on the opensource platform of Pytorch \cite{paszke2019pytorch} and all the neural networks are trained by the stochastic optimization algorithm of Adam \cite{kingma2014adam} with a constant or decaying learning rate.

\subsection{Solving Boltzmann equation for electron transport in weakly ionized plasma}
\label{sec:sec3.1}
\paragraph{}
In the kinetic modeling of plasmas, each species (e.g. electrons and heavy particles) in a plasma are treated as a collection of particles with various velocities or energies. The velocity or energy distribution of species can be obtained by solving the Boltzmann equation. In weakly ionized plasmas (e.g. the atmospheric gas discharge), the transport of electrons dominates the plasma behavior and the electron energy distribution is usually determined by solving the Boltzmann equation of electrons which has the following form in a polar coordinate \cite{hagelaar2005solving, itoh1988electron}.

\begin{equation}
\label{equ:equ18}
	\bar{R}_{i} f(v, \theta)+\frac{e E}{m}\left(\cos \theta \frac{\partial f(v, \theta)}{\partial v}-\frac{\sin \theta}{v} \frac{\partial f(v, \theta)}{\partial \theta}\right)-\left(\frac{\partial f(v, \theta)}{\partial t}\right)_{\text {coll }}=0
\end{equation}

\begin{equation}
\label{equ:equ19}
	\bar{R}_{i}=2 \pi \int_{0}^{\infty} \int_{0}^{\pi} N\left(Q_{i}(v)-Q_{a}(v)\right) v f(v, \theta) v^{2} \sin \theta d \theta d v
\end{equation}

\begin{equation}
\label{equ:equ20}
\begin{aligned}
	\left(\frac{\partial f(v, \theta)}{\partial t}\right)_{\operatorname{coll}}&=\frac{1}{2} N Q_{el}\left(v_{el}\right) v\left(1+\frac{2 m}{M}\right)^{2} \times \int_{0}^{\pi} f\left(v_{el}, \theta\right) \sin \theta d \theta\\
&+\frac{1}{2} N Q_{ex}\left(v_{ex}\right) \frac{v_{ex}^{2}}{v} \times \int_{0}^{\pi} f\left(v_{ex}, \theta\right) \sin \theta d \theta\\
&+\frac{1}{2 \Delta} N Q_{i}\left(v_{i1}\right) \frac{v_{i1}^{2}}{v} \times \int_{0}^{\pi} f\left(v_{i 1}, \theta\right) \sin \theta d \theta\\
&+\frac{1}{2(1-\Delta)} N Q_{i}\left(v_{i2}\right) \frac{v_{i2}^{2}}{v} \times \int_{0}^{\pi} f\left(v_{i2}, \theta\right) \sin \theta d \theta-N Q_{T}(v) v f(v, \theta)
\end{aligned}
\end{equation}

Where $\bar{R}_{i}$ is the effective ionization frequency; $E$ is the applied uniform electric field; $v$ is the electron velocity; $\theta$ is the polar angle with respect to the inverse direction of electric field; $e$ is the elementary charge; $m$ and $M$ are the mass of electrons and molecules respectively; $Q_{i}$, $Q_{a}$, $Q_{el}$, and $Q_{ex}$ are the electron-impact cross sections relating to the processes of ionization, electron attachment, elastic collision, and excitation; $Q_{T}$ is the total cross section which is the sum of all the other cross sections; $v_{el}$, $v_{ex}$, $v_{i1}$, and $v_{i2}$ are the electron velocities characterizing the corresponding elastic, excitation, and ionization collision processes; $\Delta$ is the ratio of the energy sharing between two electrons during ionization collision, which is usually set to 0.5 in the calculation.

\begin{figure}
	\centering
	\includegraphics[width=9.5cm]{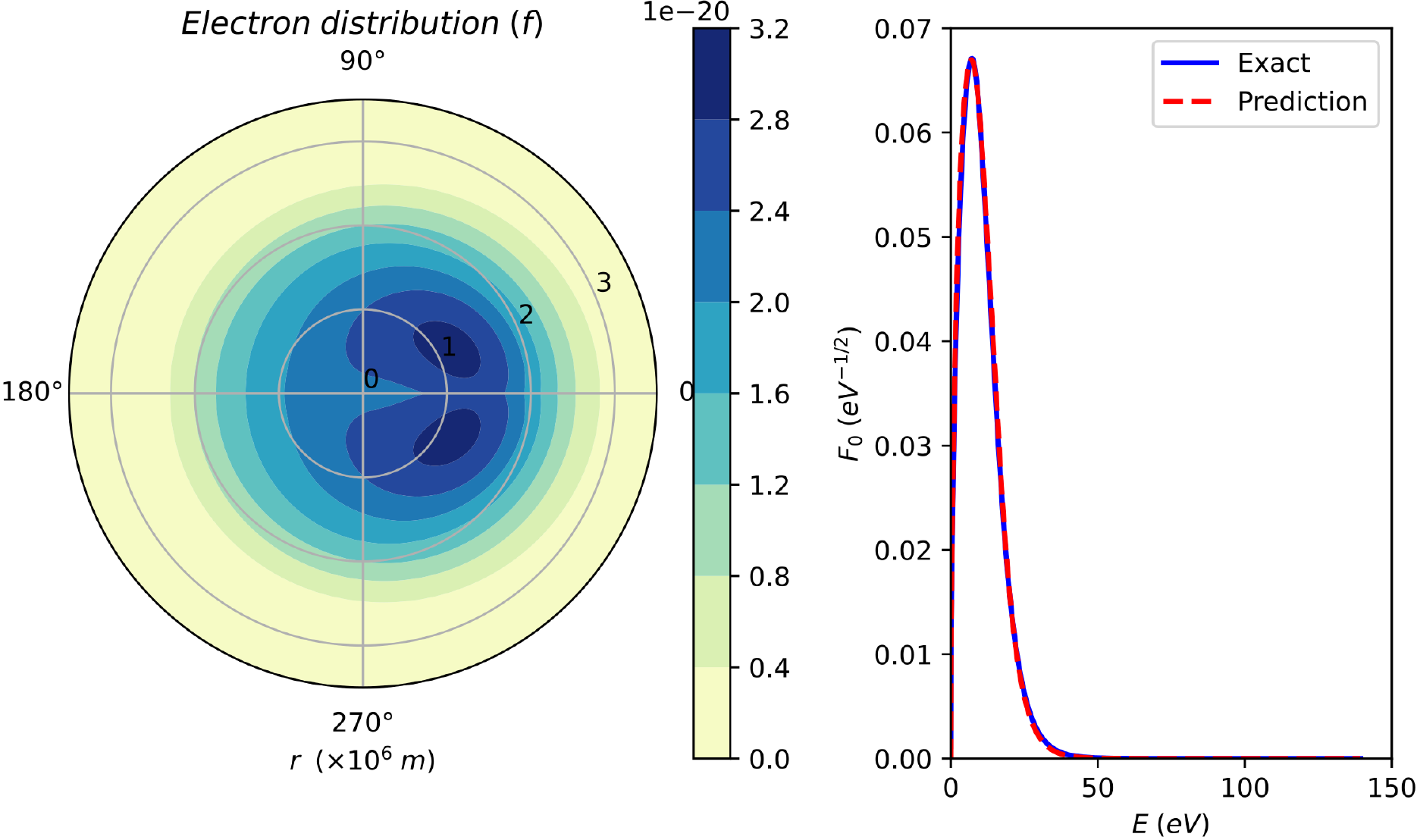}
	\caption{Electron distribution of argon plasma at E/N of 500 Td predicted by CS-PINN. The result by Bolsig+ is regarded as the exact solution. The relative $\mathrm{L}^{2}$ error for $\mathrm{F}_{0}$ is 0.023.}
	\label{fig:fig3}
\end{figure}

\paragraph{} 
In this case, we construct a neural network with 3 hidden layers and 500 neurons in each layer to map the non-linear relationship from the input $(v, \theta)$ to the solution $f$. 200 scattered points for training are sampled uniformly along the velocity and angle axis. Based on the framework of CS-PINN, we use cubic spline functions as subnet to approximate the elastic, ionization, and excitation cross sections of argon which were compiled from the website of LXCat \cite{pitchford2017lxcat}. Figure \ref{fig:fig3} illustrates the electron distribution of argon plasma at reduced electric field strength E/N of 500 Td calculated by CS-PINN after an iterative training of 20,000 epochs. The initial learning rate is set to $10^{-4}$ and decays by half every 5000 iterative steps during the training. Figure \ref{fig:fig3} also compares the electron energy distribution $\mathrm{F}_{0}$ predicted by CS-PINN with the result by Bolsig+ which is a widely used two-order Boltzmann solver \cite{hagelaar2005solving}. A good agreement can be observed in Figure \ref{fig:fig3} and the relative $\mathrm{L}^{2}$ error for $\mathrm{F}_{0}$ is 0.023.

\subsection{Solving drift-diffusion-Poisson equations for 1-D DC corona discharge plasma}
\label{sec:sec3.2}
\paragraph{}
In the plasma fluid modeling, the macroscopic quantities e.g. species densities and temperatures are calculated by solving fluid equations. To describe a corona discharge plasma which is produced by high voltage at standard atmospheric pressure, the drift-diffusion equations governing the transport of electrons and ions coupled with the Poisson equation can be used to simulate the corona behavior. The general form of the drift-diffusion-Poisson equations for corona discharge in argon can be written as follows \cite{gao2018numerical}.

\begin{equation}
\label{equ:equ21}
	\frac{\partial n_{e}}{\partial t}=\nabla \cdot\left(\mu_{e} \vec{E} n_{e}+D_{e} \cdot \nabla n_{e}\right)+\alpha n_{e}\left|\mu_{e} \vec{E}\right|
\end{equation}

\begin{equation}
\label{equ:equ22}
	\frac{\partial n_{p}}{\partial t}=\nabla \cdot\left(-\mu_{p} \vec{E} n_{p}+D_{p} \cdot \nabla n_{p}\right)+\alpha n_{e}\left|\mu_{e} \vec{E}\right|
\end{equation}

\begin{equation}
\label{equ:equ23}
	\nabla^{2} V=-\frac{e\left(n_{p}-n_{e}\right)}{\varepsilon}
\end{equation}

Where $n_{e}$, $n_{p}$, $\mu_{e}$, $\mu_{p}$, $D_{e}$, and $D_{p}$ are the number density, mobility, and diffusion coefficient of electrons and positive ions; $\alpha$ is the ionization coefficient; $E$ is the electric field strength; $V$ is the applied DC voltage; $e$ and $\varepsilon$ are the elementary charge and permittivity respectively. It is worth mentioning that argon has poor electron attachment ability. As a result, the equation of negative ions is not considered in this case and the contributions from electron attachment in the transport of electrons and positive ions are also ignored.

\paragraph{}
Since electrons transport much more quickly than heavy particles in corona discharges, the spatial distribution of the ion density is sometimes assumed to be time-independent during the short-time simulation. To simplify the modeling, the initial number densities of both electrons and positive ions in this case have the following form.

\begin{equation}
\label{equ:equ24}
	n_{e(p)}(t=0)=n_{\max }\left(k_{0}+\exp \left(-\frac{r^{2}}{2 s_{0}^{2}}\right)\right)
\end{equation}

Where $n_{max}$ is the maximum number density which is set to $10^{15} \mathrm{~m}^{-3}$ for both electrons and positive ions; $k_{0}$ and $s_{0}$ are the parameters controlling the function shape, which are 0.001 and 0.25 cm in this case. 

\paragraph{}
In the boundary, the negative DC voltage of -10 kV is applied to the cathode and the corresponding anode is grounded. When positive ions hit the surface of cathode, the secondary electrons are generated, leading to the electron flux $\Gamma_{e}$ as the boundary condition of electron equation \cite{gao2018numerical}, whereas the zero flux is assumed in the anode. The gap between two electrodes is 1.0 cm in this case.

\begin{equation}
\label{equ:equ25}
	\Gamma_{e}=\gamma n_{p} \mu_{p}|\vec{E}|
\end{equation}

Where $\gamma$ is the secondary electron emission coefficient which is set to 0.066 for corona discharges in argon.

\begin{figure}
	\centering
	\includegraphics[width=9.5cm]{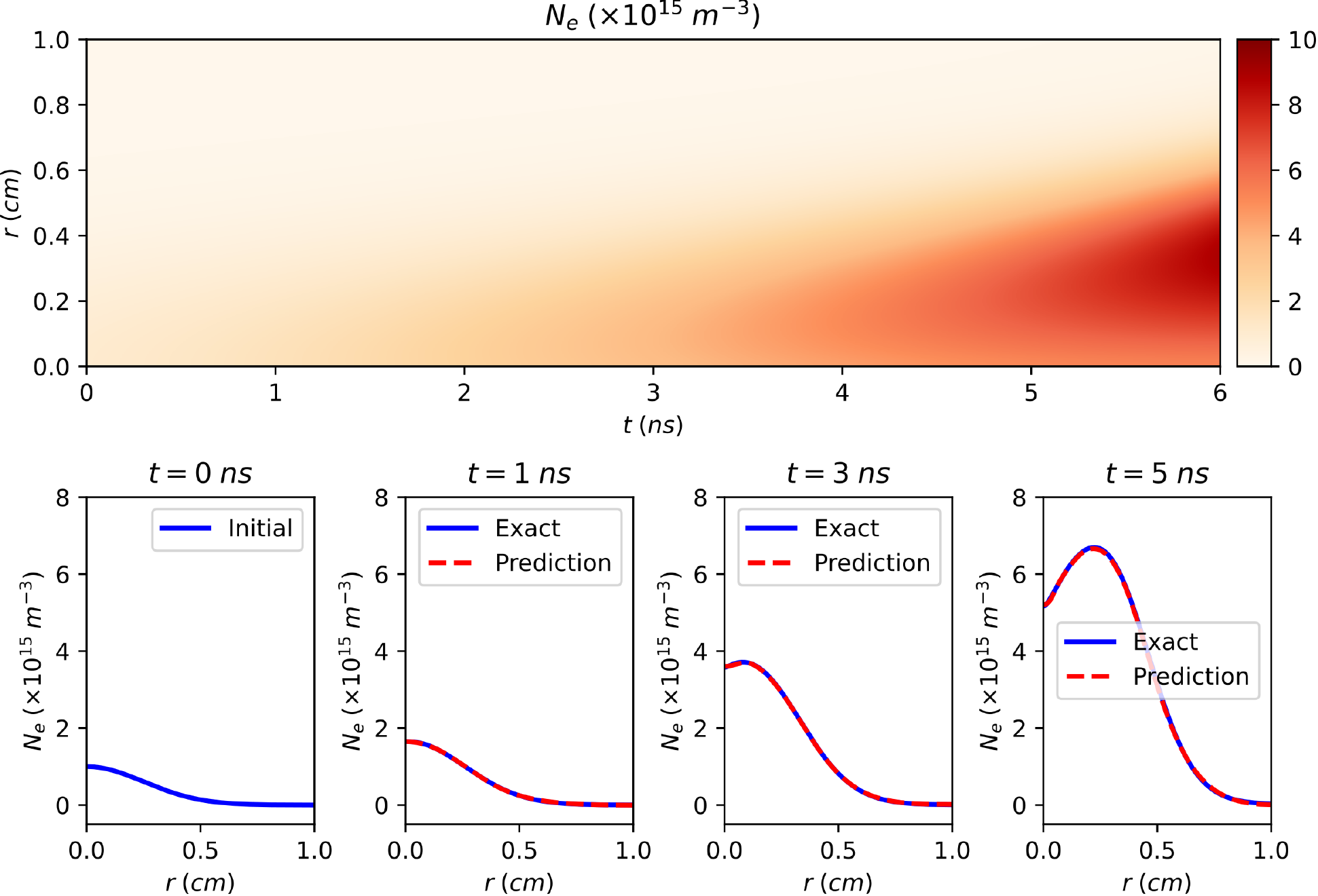}
	\caption{Number density of electrons predicted by RK-PINN. The exact solution was obtained by an explicit time-advancing method with a time step of 0.01 ns based on the opensource Chebfun package \cite{driscoll2014chebfun}. The relative $\mathrm{L}^{2}$ errors for electron density at $t$ = 1, 3, 5 ns are $6.93 \times 10^{-3}$, $4.67 \times 10^{-3}$, $8.19 \times 10^{-3}$ respectively.}
	\label{fig:fig4}
\end{figure}

\begin{figure}
	\centering
	\includegraphics[width=9.5cm]{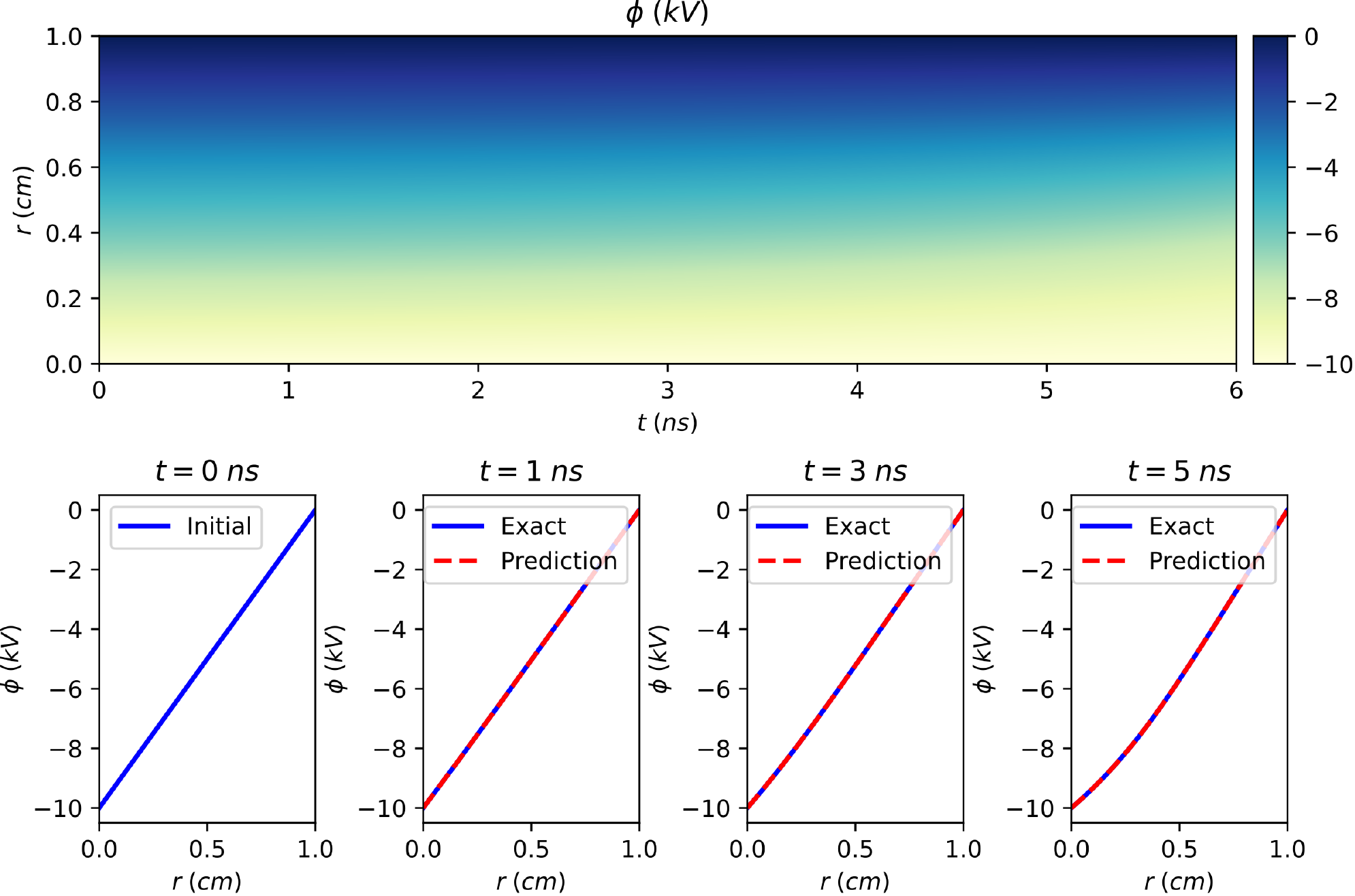}
	\caption{Electric potential predicted by RK-PINN. The exact solution was obtained by an explicit time-advancing method with a time step of 0.01 ns based on the opensource Chebfun package \cite{driscoll2014chebfun}. The relative $\mathrm{L}^{2}$ errors for electric potential at $t$ = 1, 3, 5 ns are $3.53 \times 10^{-5}$, $3.45 \times 10^{-4}$, $7.20 \times 10^{-5}$ respectively.}
	\label{fig:fig5}
\end{figure}

\paragraph{}
Due to the very fast evolution of corona discharges at high voltages, a very short time step must be selected properly in the tradition numerical methods (e.g. FEM and FVM) to ensure stability and convergence for solving corona equations. In this case, we demonstrate that the RK-PINN allows a large time step to be used in the simulation of corona plasma. We construct a feed-forward neural network with 4 hidden layers and 300 neurons in each layer to express the coupling relationship of electron and Poisson equations. The network takes spatial data as input and outputs the solution $n_{e}$ and $V$. 500 scattered points as training data are sampled uniformly along the spatial coordinate. The stage $q$ of RK-PINN is set to 300. Figures \ref{fig:fig4} and \ref{fig:fig5} compare the electron density and electric potential predicted by RK-PINN and a high-precision explicit time-advancing method. Obviously, the RK-PINN yields very good solution directly from an initial state even given a large time step. The maximum $\mathrm{L}^{2}$ errors for electron density and electric potential are $8.19 \times 10^{-3}$ and $3.45 \times 10^{-4}$ respectively. To achieve the same accuracy, the traditional explicit time-advancing method needs a very short time step of 0.01 ns while the RK-PINN can use a large time step of 1 $\sim$ 5 ns. 

\subsection{Solving time-dependent Elenbaas-Heller equation for 1-D arc plasma with radial velocity}
\label{sec:sec3.3}
\paragraph{}
The fluid approach can also be applied to simulate arc plasmas which are widely used as energy source in industry. To describe the behavior of arc plasmas, the fluid equations governing mass continuity, momentum conservation, and energy conservation are coupled with Maxwell’s equations, which is also referred as MHD equations. In this case, a 1-D transient arc with radial velocity is studied based on CS-PINN and RK-PINN. The governing equations of radial velocity $\nu_{r}$ and arc temperature $T$ are written as follows \cite{zhong2019animproved}, which is a time-dependent variant of Elenbaas-Heller equation.

\begin{equation}
\label{equ:equ26}
	\frac{\partial \rho}{\partial t}+\frac{1}{r} \frac{\partial}{\partial r}\left(r \rho v_{r}\right)=0
\end{equation}

\begin{equation}
\label{equ:equ27}
	\rho C_{p}\left(\frac{\partial T}{\partial t}+v_{r} \frac{\partial T}{\partial r}\right)=\frac{1}{r} \frac{\partial}{\partial r}\left(r \kappa \frac{\partial T}{\partial r}\right)+\sigma \frac{I^{2}}{g^{2}}-4 \pi \varepsilon
\end{equation}

Where $\rho$, $C_{p}$, $\kappa$, $\sigma$, and $\varepsilon$ are the mass density, specific heat, thermal conductivity, electrical conductivity, and net emission coefficient of an arc plasma, respectively; $I$ is the current;$ \mathbf{g}$ is the arc conductance which is calculated by the following integral within the arc radius $R$ \cite{zhong2019evaluation}.

\begin{equation}
\label{equ:equ28}
	g=\int_{0}^{R} 2 \pi r \sigma d r
\end{equation}

\paragraph{}
The initial velocity is zero and the initial temperature is determined by the stationary arc model \cite{zhong2019animproved, zhong2019evaluation}. The initial current is 200 A and set to be zero after that ($t>0$) \cite{zhong2020deep}. At the symmetry ($r = 0$), the radial velocity and temperature gradient are assumed to be zero. At the boundary wall, the temperature is set to be constant (2000 K in this case).

\begin{figure}
	\centering
	\includegraphics[width=9.5cm]{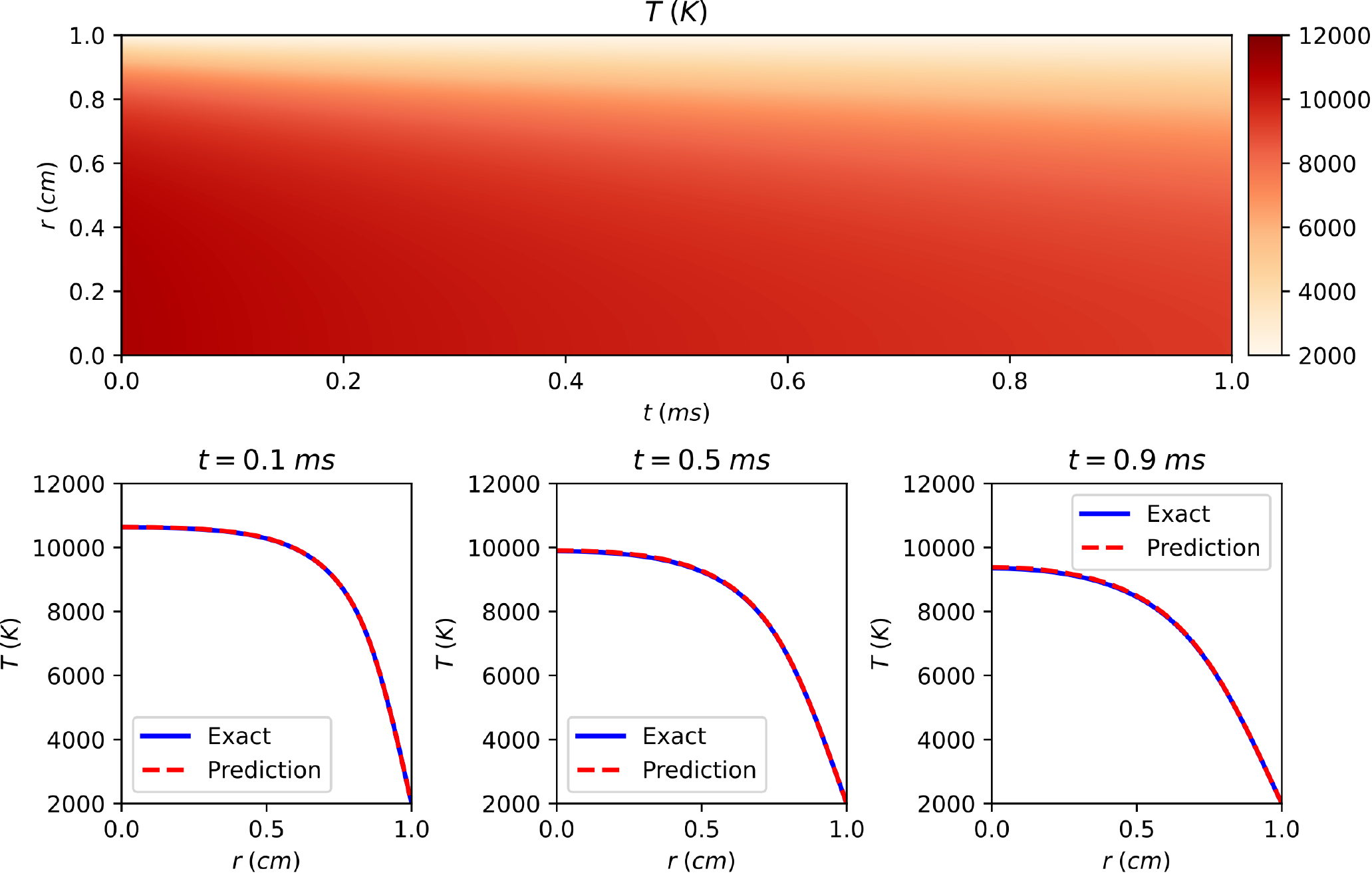}
	\caption{Spatial and temporal distribution of arc temperature in argon plasma at ambient pressure predicted by CS-PINN. The exact solution was obtained by the explicit time-advancing method with a time step of 1 ns. The relative $\mathrm{L}^{2}$ errors for temperature distribution at $t$ = 0.1, 0.5, 0.9 ms are $7.77 \times 10^{-4}$, $2.36 \times 10^{-3}$, $3.34 \times 10^{-3}$ respectively.}
	\label{fig:fig6}
\end{figure}

\begin{figure}
	\centering
	\includegraphics[width=9.5cm]{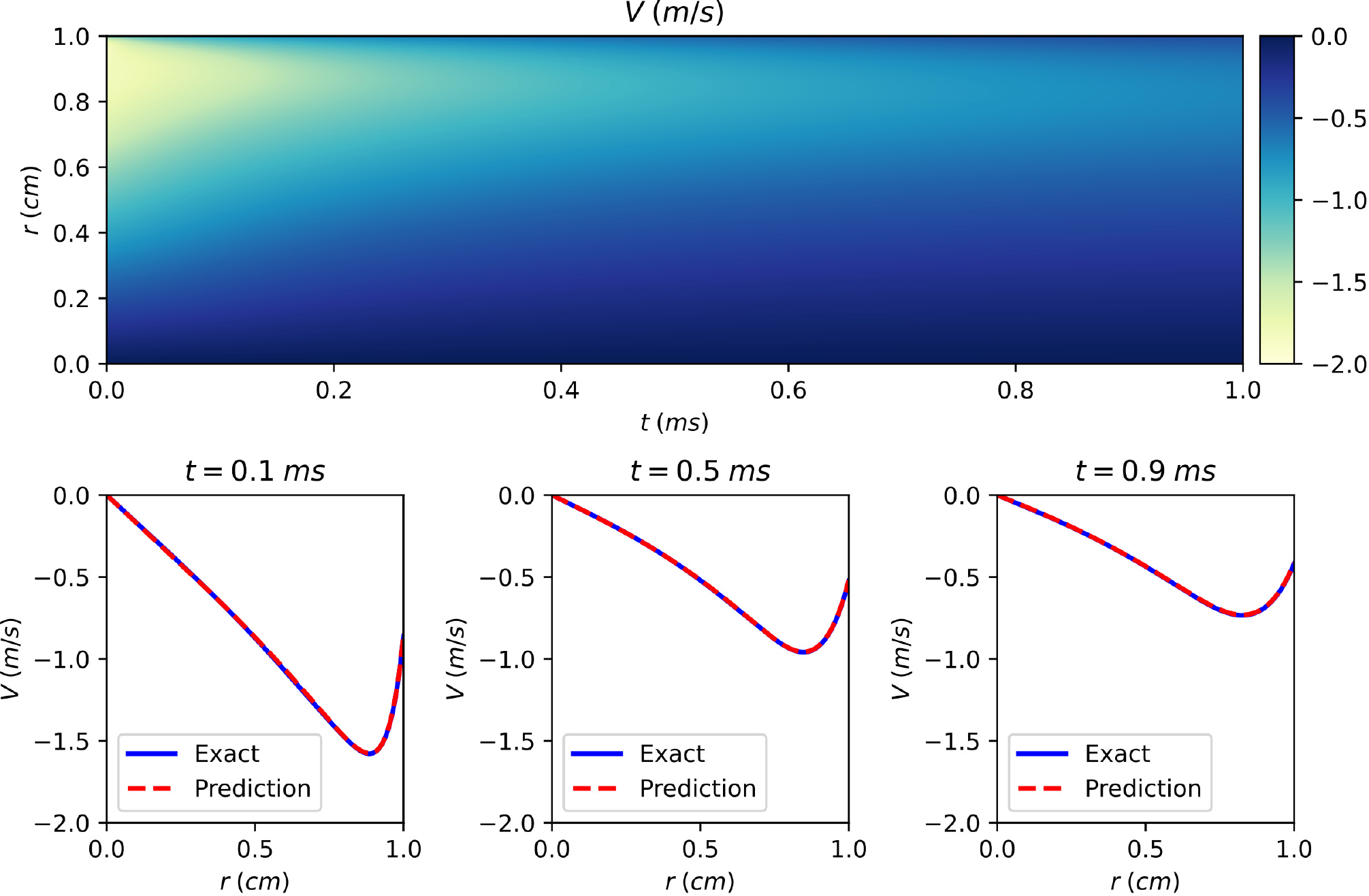}
	\caption{Spatial and temporal distribution of radial velocity in argon plasma at ambient pressure predicted by CS-PINN. The exact solution was obtained by the explicit time-advancing method with a time step of 1 ns.  The relative $\mathrm{L}^{2}$ errors for velocity distribution at $t$ = 0.1, 0.5, 0.9 ms are $4.09 \times 10^{-3}$, $2.90 \times 10^{-3}$, $4.82 \times 10^{-3}$ respectively.}
	\label{fig:fig7}
\end{figure}

\paragraph{}
In the framework of CS-PINN, we construct a neural network with 6 hidden layers and 300 neurons in each layer to surrogate the 1-D arc equation of argon plasma at ambient pressure. The network takes $(t, r)$ as input and outputs the solutions $v_{r}$ and $T$. All the thermodynamic and transport coefficients (e.g. $\rho$, $C_{p}$, $\kappa$, $\sigma$, and $\varepsilon$) of argon plasma in the arc equation are approximated by cubic spline functions as subnet in CS-PINN. To prepare the training data, 200 and 100 scattered points are sampled uniformly along the radial axis and timeline respectively. The arc radius is set to 1 cm in this case. The spatial and temporal distributions of arc temperature and radial velocity predicted by CS-PINN are described in Figures \ref{fig:fig6} and \ref{fig:fig7} respectively. The CS-PINN prediction is also compared with the results by an explicit time-advancing method with a short time step of 1 ns. The maximum $\mathrm{L}^{2}$ relative errors for arc temperature and radial velocity are $1.31 \times 10^{-3}$ and $6.07 \times 10^{-3}$ respectively.

\begin{figure}
	\centering
	\includegraphics[width=9.5cm]{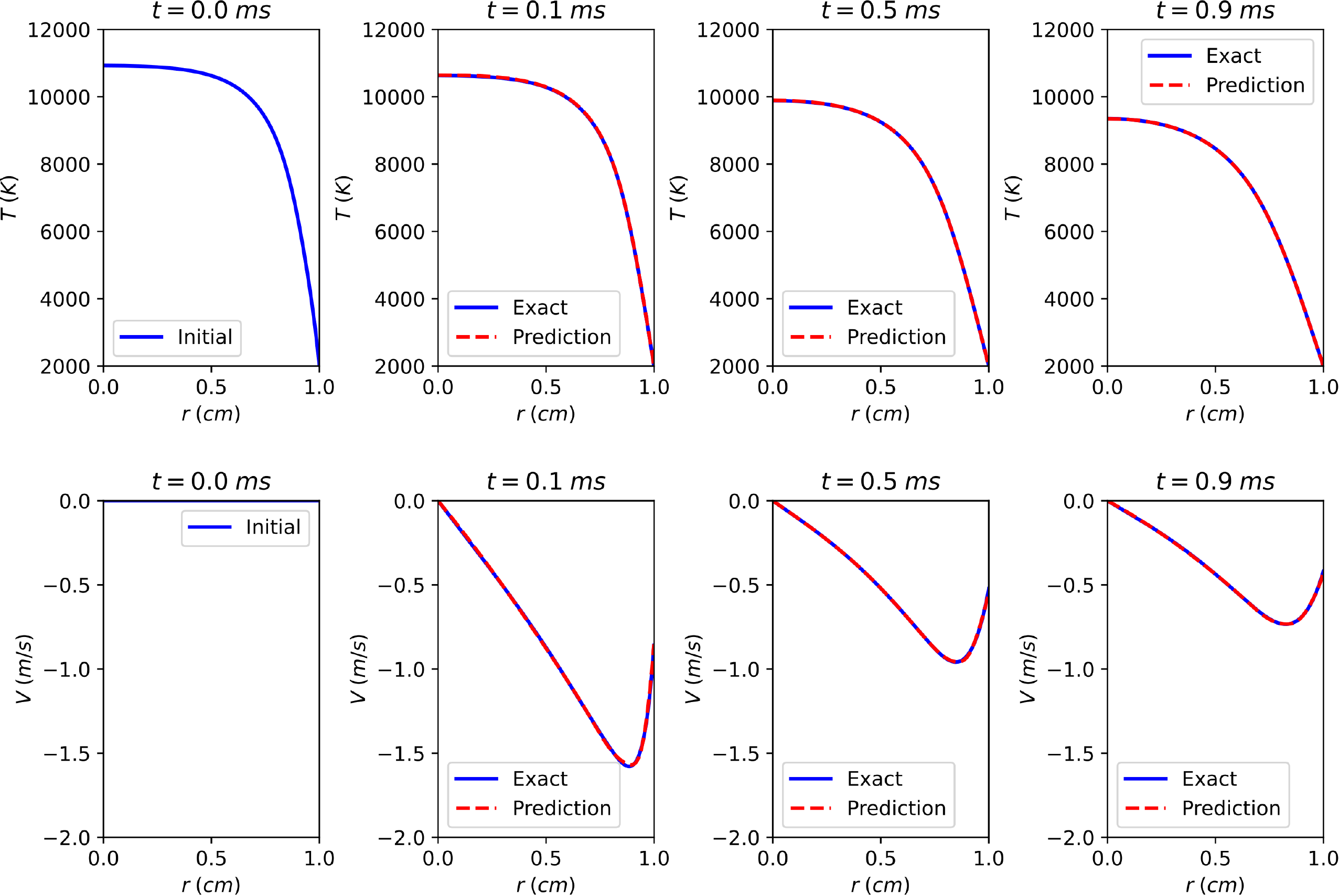}
	\caption{Arc temperature and radial velocity of in argon plasma at ambient pressure predicted by the RK-PINN. The exact solution was obtained by the explicit time advancing method with a time step of 1 ns. The relative $\mathrm{L}^{2}$ errors for arc temperature at $t$ = 0.1, 0.5, 0.9 ms are $1.31 \times 10^{-3}$, $5.87 \times 10^{-4}$, $2.59 \times 10^{-4}$ respectively. The relative $\mathrm{L}^{2}$ errors for radial velocity at $t$ = 0.1, 0.5, 0.9 ms are $6.07 \times 10^{-3}$, $5.83 \times 10^{-3}$, $4.26 \times 10^{-3}$ respectively.}
	\label{fig:fig8}
\end{figure}

\paragraph{}
	We also demonstrate that the time-dependent Elenbaas-Heller equation for 1-D arc plasma can be easily solved by RK-PINN. In the RK-PINN framework, we construct a neural network with 4 hidden layers and 500 neurons in each layer to learn the coupling relationship of continuity equation  and energy conservation equation. 500 scattered points are sampled as training data uniformly along the radial axis. The stage $q$ in the implicit Runge-Kutta formalism is set to 300. The arc temperature and radial velocity predicted by RK-PINN directly from $t$ = 0.0 to 0.1, 0.5, and 0.9 ms are compared in Figure \ref{fig:fig8} with the results by a high-precision explicit time-advancing method. The maximum $\mathrm{L}^{2}$ errors for arc temperature and radial velocity are $1.31 \times 10^{-3}$ and $6.07 \times 10^{-3}$ respectively, which are on the same order of magnitude as the results by CS-PINN. Compared with the traditional explicit time-advancing method which needs a very short time step of 1 ns, the RK-PINN can use a large time step of 0.1 $\sim$ 0.9 ms.

\subsection{Solving time-dependent Elenbaas-Heller equation for 1-D arc plasma without initial condition}
\label{sec:sec3.4}
\paragraph{}
It is known that numerical solving of PDEs requires a certain initial and/or boundary conditions to yield a fixed solution. However, it is sometimes kind of difficult in practical applications to decide an exact initial and boundary conditions. For instance, in a transient arc discharge, it is not easy to exactly identify which initial state the arc plasma starts from because the arc is evolving very fast. Consequently, it is very necessary to develop a method which can solve PDEs without initial and/or boundary conditions. As a data-driven method in essence, PINNs provide us an approach to doing this.

\begin{figure}
	\centering
	\includegraphics[width=9.5cm]{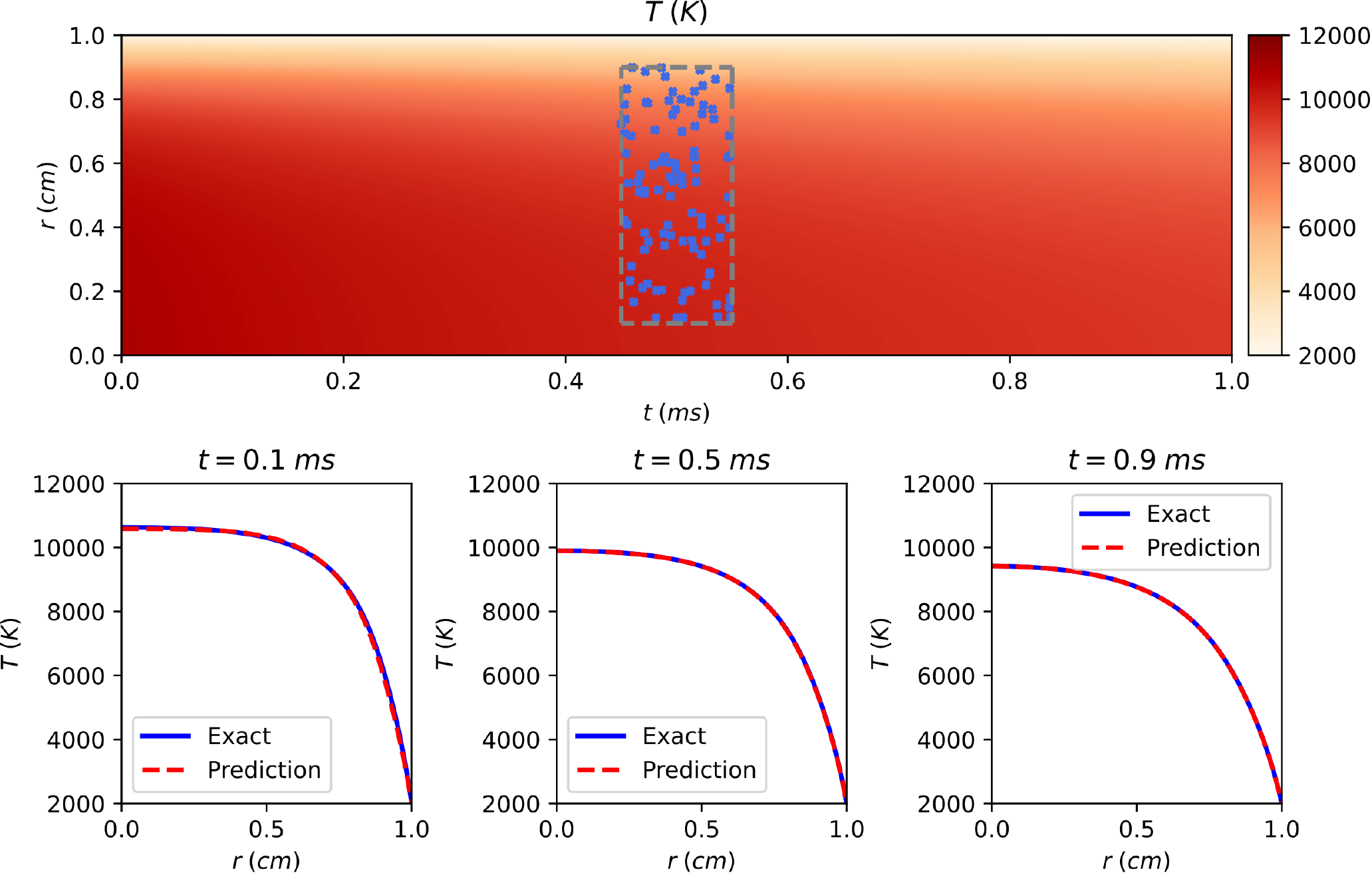}
	\caption{Spatial and temporal distribution of arc temperature in argon plasma at ambient pressure predicted by CS-PINN without an initial condition. The exact solution was obtained by an explicit time-advancing method with a time step of 1 ns. The relative $\mathrm{L}^{2}$ errors for temperature distribution at $t$ = 0.1, 0.5, 0.9 ms are $7.75 \times 10^{-3}$, $1.54 \times 10^{-3}$, $8.52 \times 10^{-4}$ respectively.}
	\label{fig:fig9}
\end{figure}

\paragraph{}
In this case, we demonstrate that how CS-PINN and RK-PINN can be modified to deal with the plasma equations without initial conditions. To solve PDEs without defined conditions, extra information e.g. sensing data in the computational domain must be supplemented. As shown in Figure \ref{fig:fig9}, we collect some sensing data of temperature (denoted as blue dots in the dotted rectangle) in a small spatio-temporal domain. These sensing data are used as extra information to make up for the lost information resulting from undefined initial condition. Meanwhile, the loss function has to be modified slightly as follows in the framework of CS-PINN.

\begin{equation}
\label{equ:equ29}
	\mathcal{L}=\omega_{f} \mathcal{L}_{f}+\omega_{\mathcal{B}} \mathcal{L}_{\mathcal{B}}+\omega_{\mathcal{D}} \mathcal{L}_{\mathcal{D}}
\end{equation}

\begin{equation}
\label{equ:equ30}
	\mathcal{L}_{\mathcal{D}}=\frac{1}{N_{\mathcal{D}}} \sum_{i=1}^{N_{\mathcal{D}}}\left\|u_{i}-\hat{u}_{i}\right\|
\end{equation}

Where $\mathcal{L}_{\mathcal{D}}$ is the loss term corresponding to the sensing data; $\omega_{\mathcal{D}}$ is the weighting factor; $N_{\mathcal{D}}$ is the number of sensing data in the given domain. 

\paragraph{}
As an example, we solve the time-dependent Elenbaas-Heller equation without radial velocity \cite{zhong2020deep}. A neural network with 6 hidden layers and 200 neurons in each layer is constructed which takes $(t, \mathbf{x})$ as input and outputs the solution $T$. As in the case C, 200 and 100 scattered points are sampled as training data uniformly along the radial axis and timeline respectively. Besides, we randomly select 100 sensing data points locating from 0.1 $\sim$ 0.9 cm during the period of 0.45 $\sim$ 0.55 ms. The result in Figure \ref{fig:fig9} indicates that even without an initial condition, the arc temperature of argon plasma can be determined from the arc equation by CS-PINN as long as some sensing data are provided. The relative $\mathrm{L}^{2}$ errors for the predicted temperature distribution at $t$ = 0.1, 0.5, and 0.9 ms are $7.75 \times 10^{-3}$, $1.54 \times 10^{-3}$, and $8.52 \times 10^{-4}$ respectively, showing that a well agreement and good accuracy has been reached.

\begin{figure}
	\centering
	\includegraphics[width=9.5cm]{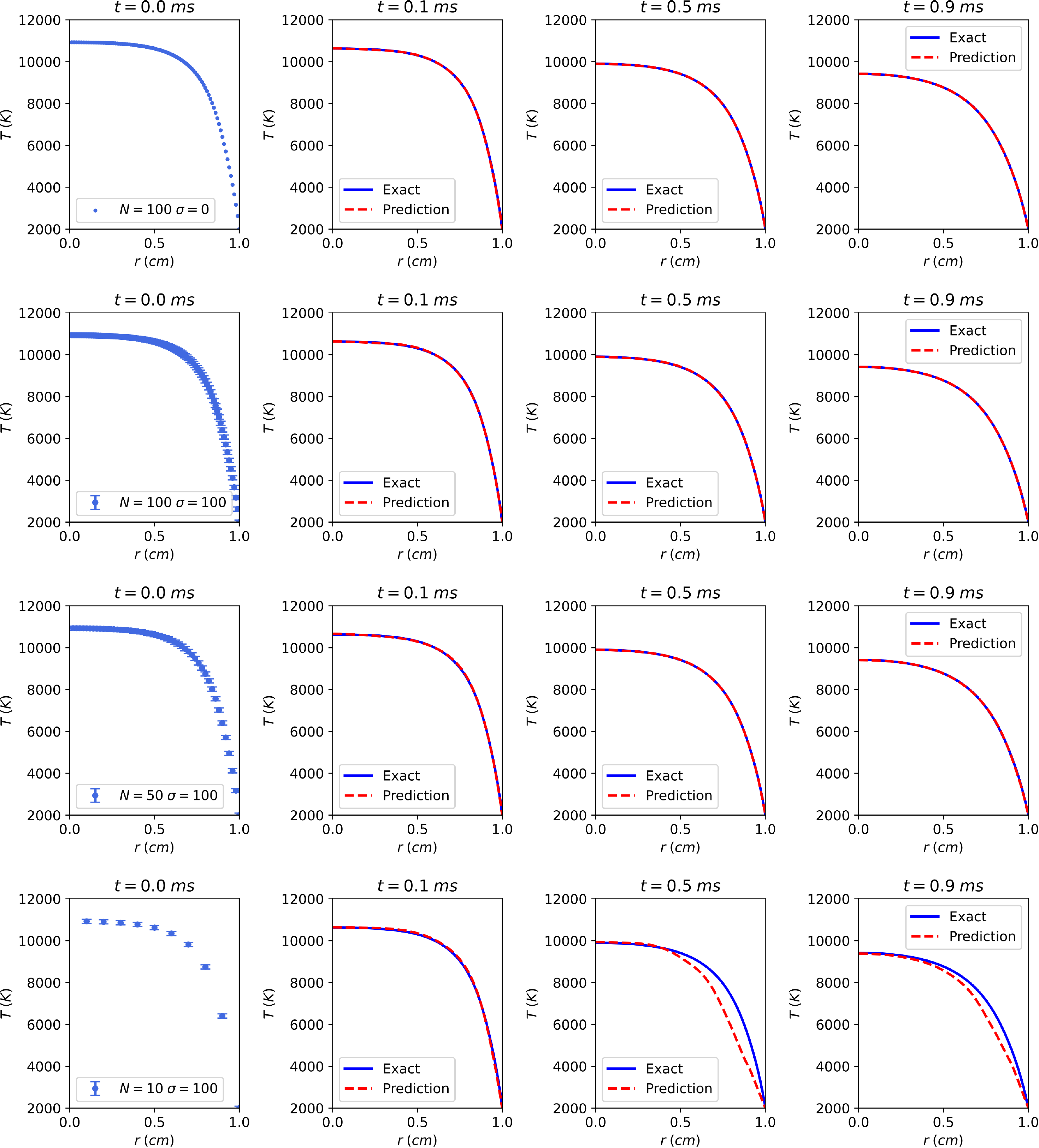}
	\caption{Spatial distribution of arc temperature in argon plasma at ambient pressure predicted by RK-PINN without an initial condition. The exact solution was obtained by an explicit time advancing method with a time step of 1 ns. The relative $\mathrm{L}^{2}$ errors for the prediction are listed in Table \ref{tab:tab1}.}
	\label{fig:fig10}
\end{figure}

\paragraph{}
In the framework of RK-PINN, we can also use the sensing data to bridge the gap causing by undefined initial conditions. It is more natural for RK-PINN than CS-PINN to do this. The corresponding loss functions are not even needed to be modified. We construct a neural network with 4 hidden layers and 300 neurons in each layer. The stage q in the implicit Runge-Kutta formalism is set to 200. As shown in Figure \ref{fig:fig10}, we firstly select 100 sensing data uniformly located along the radial coordinate and find that the temperature distributions at $t$ = 0.1, 0.5, and 0.9 ms can be accurately predicted by RK-PINN. The corresponding relative $\mathrm{L}^{2}$ errors are listed in Table \ref{tab:tab1}. Next, we add some Gaussian noises with a standard deviation $\sigma=100 \mathrm{~K}$ in the sensing data. The result reveals that the RK-PINN also outputs good predictions despite noises in the sensing data. Interestingly, when we keep the same noise level and reduce the number of sensing data by half, the relative $\mathrm{L}^{2}$ errors of predicted temperature distribution by RK-PINN do not decline significantly. However, when we further reduce the number of sensing data from 50 to 10, only a small time step $(\Delta t=0.1 \mathrm{~ms})$ can ensure a good prediction by RK-PINN. Anyway, compared with the previous prediction using 50 sensing data, the large time steps $(\Delta t=0.5$ and $0.9 \mathrm{~ms})$ with only 10 sensing data increases the $\mathrm{L}^{2}$ error by only one order of magnitude on average. This is acceptable especially when collecting sensing data is costly.

\begin{table}[htbp]
	\centering
	\caption{The relative $\mathrm{L}^{2}$ errors for temperature distribution predicted by RK-PINN using different sensing data at $t$ = 0.1, 0.5, 0.9 ms}
	\renewcommand{\arraystretch}{0.5}
	\begin{tabular}{cccc}
		\label{tab:tab1}\\
		\toprule 
		&\\&0.1ms&0.5ms&0.9ms\\ 
		\midrule
		$N=100, \sigma=0$ & $2.17 \times 10^{-3}$ & $6.23 \times 10^{-4}$ & $6.58 \times 10^{-4}$ \\
		$N=100, \sigma=100$ & $1.83 \times 10^{-3}$ & $1.21 \times 10^{-3}$ & $1.05 \times 10^{-3}$ \\
		$N=50, \sigma=100$ & $2.68 \times 10^{-3}$ & $8.23 \times 10^{-4}$ & $1.92 \times 10^{-3}$ \\
		$N=10, \sigma=100$ & $6.74 \times 10^{-3}$ & $8.41 \times 10^{-2}$ & $5.21 \times 10^{-2}$ \\
		\bottomrule 
	\end{tabular}
\end{table}

\section{CONCLUSIONS}
\label{sec:sec4}
\paragraph{}
In this work, we propose two general AI-driven frameworks for low-temperature plasma simulation: Coefficient-Subnet Physics-Informed Neural Network (CS-PINN) and Runge-Kutta Physics-Informed Neural Network (RK-PINN). Both CS-PINN and RK-PINN leverage the powerful nonlinear expressiveness of neural networks to learn the mapping relationship between spatio-temporal space and equation’s solution. To effectively approximate solution-dependent coefficients (e.g. electron-impact cross sections, thermodynamic properties, transport coefficients, et al.), the CS-PINN use either a neural network or an interpolation function (e.g. spline function) as the subnet of the framework. Based on the CS-PINN, the RK-PINN incorporates the implicit Runge-Kutta formalism in neural networks and their corresponding loss functions, which realizes a fast prediction of plasma equations with transient terms. We demonstrate preliminary applications of these two frameworks by solving different plasma governing equations, including the Boltzmann equation for electron transport in weakly ionized plasma, the drift-diffusion-Poisson equations for 1-D DC corona discharge plasma, the time-dependent Elenbaas-Heller equation for 1-D arc plasma with radial velocity, and the time-dependent Elenbaas-Heller equation for 1-D arc plasma without initial condition. The results show that both CS-PINN and RK-PINN perform well in solving plasma equations. Particularly, the RK-PINN can yield a good solution not only with large time step but also using limited noisy sensing data. All of these indicate a promising new tool for plasma simulation. Still, there are much work to do in the future, such as improving the efficiency and convergence of training neural networks for solving large-scale multi-physics equations.

\section*{Acknowledgments}
\label{sec:acknowledgments}
\paragraph{}
This work was supported in part by the National Natural Science Foundation of China (92066106, 51907023), the Young Scientific and Technical Talents Promotion Project of Jiangsu Association for Science and Technology (2021031), the Zhishan Young Scholar Project of Southeast University, and the Fundamental Research Funds for the Central Universities.

\bibliographystyle{unsrt}
%\bibliography{references}  %%% Remove comment to use the external .bib file (using bibtex).
%%% and comment out the ``thebibliography'' section.

%%% Comment out this section when you \bibliography{references} is enabled.

\begin{thebibliography}{10}

	\bibitem{zhang2018amoo3}
	G.~Zhang, T.~Xiong, M.~Yan, L.~He, X.~Liao, C.~He, C.~Yin, H.~Zhang, and L.~Mai.
	\newblock $\alpha$-MoO3-x by plasma etching with improved capacity and stabilized structure for lithium storage.
	\newblock {\em Nano Energy}, 49:555--563, 2018.
	
	\bibitem{han2018recent}
	Z.~J.~Han, A.~T.~Murdock, D.~H.~Seo, and A.~Bendavid.
	\newblock Recent progress in plasma-assisted synthesis and modification of 2D materials.
	\newblock {\em 2D Materials}, 5:032002, 2018.
	
	\bibitem{zhong2019animproved}
	L.~Zhong, Q.~Gu, and S.~Zheng.
	\newblock An improved method for fast evaluating arc quenching performance of a gas based on 1D arc decaying 
	model.
	\newblock {\em Phys. Plasmas}, 26(10):103507, 2019.	

	\bibitem{li2021advances}
	M.~Li, Z.~Wang, R.~Xu, X.~Zhang, Z.~Chen, and Q.~Wang.
	\newblock Advances in plasma-assisted ignition and combustion for combustors of aerospace engines.
	\newblock {\em Aerospace Science and Technology}, 117:106952, 2021.
	
	\bibitem{Levchenko2020perspectives}
	I.~Levchenko, S.~Xu, S.~Mazouffre, D.~Lev, D.~Pedrini, D.~Goebel, L.~Garrigues, F.~Taccogna, and K.~Bazaka.
	\newblock Perspectives, frontiers, and new horizons for plasma-based space electric propulsion.
	\newblock {\em Phys. Plasmas}, 27(2):020601, 2020.

	\bibitem{sanito2021application}
	R.~C.~Sanito, S.~-J. You, and Y.~-F. Wang.
	\newblock Application of plasma technology for treating e-waste: A review.
	\newblock {\em Journal of environmental management}, 288:112380, 2021.

	\bibitem{cheng2020onthe}
	H.~Cheng, J.~Xu, X.~Li, D.~Liu, and X.~Lu.
	\newblock On the dose of plasma medicine: Equivalent total oxidation potential (ETOP).
	\newblock {\em Phys. Plasmas}, 27(6):063514, 2020.

	\bibitem{loureiro2016kinetics}
	J.~Loureiro, and J.~Amorim.
	\newblock Kinetics and spectroscopy of low temperature plasmas.
	\newblock {\em Springer International Publishing}, 2016.

	\bibitem{zhong2021dynamics}
	H.~Zhong, M.~N. Shneider, X.~Mao, and Y.~Ju.
	\newblock Dynamics and chemical mode analysis of plasma thermal-chemical instability.
	\newblock {\em Plasma Sources Science and Technology}, 30(3):035002, 2021.

	\bibitem{gleizes2014perspectives}
	A.~Gleizes.
	\newblock Perspectives on Thermal Plasma Modelling.
	\newblock {\em Plasma Chemistry and Plasma Processing}, 35(3):455-469, 2014.

	\bibitem{kochkov2021machine}
	D.~Kochkov, J.~A. Smith, A.~Alieva, Q.~Wang, M.~P.~Brenner, and S.~Hoyer.
	\newblock Machine learning accelerated computational fluid dynamics.
	\newblock {\em arXiv preprint arXiv:2102.01010}, 2021.

	\bibitem{raissi2019physics}
	M.~Raissi, P.~Perdikaris, and G.~E. Karniadakis.
	\newblock Physics-informed neural networks: A deep learning framework for solving forward and inverse problems 			involving nonlinear partial differential equations.
	\newblock {\em Journal of Computational Physics}, 378:686--707, 2019.

	\bibitem{cai2021physicsfluid}
	S.~Cai, Z.~Mao, Z.~Wang, M.~Yin, and G.~E. Karniadakis.
	\newblock Physics-informed neural networks (PINNs) for fluid mechanics: A review.
	\newblock {\em arXiv preprint arXiv:2105.09506}, 2021.

	\bibitem{pang2019fpinns}
	G.~Pang, L.~Lu, and G.~E. Karniadakis.
	\newblock fPINNs: Fractional physics-informed neural networks.
	\newblock {\em SIAM Journal on Scientific Computing}, 41(4):A2603--A2626, 2019.

	\bibitem{jagtap2020extended}
	A.~D.~Jagtap, and G.~E. Karniadakis.
	\newblock Extended physics-informed neural networks (xpinns): A generalized space-time domain decomposition 			based deep learning framework for nonlinear partial differential equations.
	\newblock {\em Communications in Computational Physics}, 28:2002, 2020.

	\bibitem{chen2021learning}
	X.~Chen, J.~Duan, and G.~E. Karniadaki.
	\newblock Learning and meta-learning of stochastic advection–diffusion–reaction systems from sparse 					measurements.
	\newblock {\em European Journal of Applied Mathematics}, 32(3):397--420, 2021.
	
	\bibitem{yang2019adversarial}
	Y.~Yang, and P.~Perdikaris.
	\newblock Adversarial uncertainty quantification in physics-informed neural networks.
	\newblock {\em Journal of Computational Physics}, 394:136--152, 2019.

	\bibitem{rodriguez2022physics}
	R.~Rodriguez-Torrado, P.~Ruiz, L.~Cueto-Felgueroso, M.~C. Green, T.~Friesen, S.~Matringe, and J.~Togelius.
	\newblock Physics-informed attention-based neural network for hyperbolic partial differential equations: 				application to the Buckley–Leverett problem.
	\newblock {\em Scientific reports}, 12(1):1--12, 2022.
	
	\bibitem{jagtap2020adptive}
	A.~D.~Jagtap, K.~Kawaguchi, and G.~E. Karniadakis.
	\newblock Adaptive activation functions accelerate convergence in deep and physics-informed neural networks.
	\newblock {\em  Journal of Computational Physics}, 404:109136, 2020.

	\bibitem{raissi2020hidden}
	M.~Raissi, A.~Yazdani, and G.~E. Karniadakis.
	\newblock Hidden fluid mechanics: Learning velocity and pressure fields from flow visualizations.
	\newblock {\em Science}, 367(6481):1026--1030, 2020.

	\bibitem{cai2021physicsheat}
	S.~Cai, Z.~Wang, S.~Wang, P.~Perdikaris, and G.~Karniadakis.
	\newblock Physics-Informed Neural Networks (PINNs) for Heat Transfer Problems.
	\newblock {\em Journal of Heat Transfer}, 143:102719, 2021.

	\bibitem{pun2019physically}
	G.~P.~P.~Pun, R.~Batra, R.~Ramprasad, and Y.~Mishin.
	\newblock Physically informed artificial neural networks for atomistic modeling of materials.
	\newblock {\em Nature Communications}, 10:2339, 2019.

	\bibitem{ji2021stiff}
	W.~Ji, W.~Qiu, Z.~Shi, S.~Pan, and S.~Deng.
	\newblock Stiff-pinn: Physics-informed neural network for stiff chemical kinetics.
	\newblock {\em The Journal of Physical Chemistry A}, 125(36):8098--8106, 2021.

	\bibitem{kawaguchi2020deep}
	S.~Kawaguchi, K.~Takahashi, H.~Ohkama, and K.~Satoh.
	\newblock Deep learning for solving the Boltzmann equation of electrons in weakly ionized plasma.
	\newblock {\em Plasma Sources Science and Technology}, 29(2):025021, 2020.

	\bibitem{zhong2020deep}
	L.~Zhong, Q.~Gu, andB.~Wu.
	\newblock Deep learning for thermal plasma simulation: Solving 1-D arc model as an examplea.
	\newblock {\em Computer Physics Communications}, 257:107496, 2020.

	\bibitem{hagelaar2005solving}
	G.~J.~M.~Hagelaar, and L.~C.~Pitchford.
	\newblock Solving the Boltzmann equation to obtain electron transport coefficients and rate coefficients for 			fluid models.
	\newblock {\em Plasma Sources Science and Technology}, 14(4):722, 2005.

	\bibitem{gao2018numerical}
	Q.~Gao, C.~Niu, K.~Adamiak, A.~Yang, M.~Rong, and X.~Wang.
	\newblock Numerical simulation of negative point-plane corona discharge mechanism in SF6 gas.
	\newblock {\em Plasma Sources Science and Technology}, 27(11):115001, 2018.

	\bibitem{paszke2019pytorch}
	A.~Paszke, S.~Gross, F.~Massa, A.~Lerer, J.~Bradbury, G.~Chanan, T.~Killeen, Z.~Lin, N.~Gimelshein,
 	and L.~Antiga.
	\newblock PyTorch: An imperative style, high-performance deep learning library.
	\newblock {\em Vancouver, Canada}, 2019.

	\bibitem{kingma2014adam}
	D.~P. Kingma, and J.~Ba.
	\newblock Adam: A method for stochastic optimization.
	\newblock {\em arXiv preprint arXiv:1412.6980}, 2014.

	\bibitem{itoh1988electron}
	H.~Itoh, Y.~Miura, N.~Ikuta, Y.~Nakao, and H.~Tagashira.
	\newblock Electron swarm development in SF6. I. Boltzmann equation analysis.
	\newblock {\em Journal of Physics D: Applied Physics}, 21(6):922, 1988.

	\bibitem{pitchford2017lxcat}
	L.~C. Pitchford, L.~L. Alves, K.~Bartschat, S.~F. Biagi, M.-C. Bordage, I.~Bray, C.~E. Brion, M.~J. Brunger, 
	L.~Campbell, A.~Chachereau, B.~Chaudhury, L.~G. Christophorou, E.~Carbone, N.~A. Dyatko, C.~M. Franck, 
	D.~V. Fursa, R.~K. Gangwar, V.~Guerra, P.~Haefliger, G.~J.~M. Hagelaar, A.~Hoesl, Y.~Itikawa, I.~V. Kochetov,
	R.~P. McEachran, W.~L. Morgan, A.~P. Napartovich, V.~Puech, M.~Rabie, L.~Sharma, R.~Srivastava, A.~D. Stauffer,
	J.~Tennyson, J.~de Urquijo, J.~van Dijk, L.~A. Viehland, M.~C. Zammit, O.~Zatsarinny, and S.~Pancheshnyi.
	\newblock LXCat: an Open-Access, Web-Based Platform for Data Needed for Modeling Low Temperature Plasmas.
	\newblock {\em Plasma Processes and Polymers}, 14(1--2):1600098, 2017.

	\bibitem{driscoll2014chebfun}
	T.~A. Driscoll, N.~Hale, and L.~N. Trefethen.
	\newblock Chebfun guide.
	\newblock {\em Pafnuty Publications, Oxford}, 2014.


	\bibitem{zhong2019evaluation}
	 L.~Zhong, Y.~Cressault, and P.~Teulet.
	\newblock Evaluation of Arc Quenching Ability for a Gas by Combining 1-D Hydrokinetic Modeling and Boltzmann 			Equation Analysis.
	\newblock {\em IEEE Trans. Plasma Sci}, 47(4):1835--1840, 2019.
	
\end{thebibliography}

\end{document}